\documentstyle[aps,prl,preprint,floats,epsfig]{revtex}

\begin{document}

\preprint{\tighten\vbox{\hbox{\hfil CLNS 98/1574}
                        \hbox{\hfil CLEO 98-11}
}}

\newcommand{\GG}{\mbox{$\gamma\gamma$}}
\newcommand{\KPZ}{\mbox{$K^{*-} \to K^- \pi^0$}}
\newcommand{\KSP}{\mbox{$K^{*-} \to K_S \pi^- \to \pi^- \pi^+ \pi^-$}}
\newcommand{\KSPP}{\mbox{$K^{*-} \to K_S \pi^-$}}

\title{\large \bf First Observation of the Decay $\tau^-
\to K^{*-}\eta \nu_{\tau}$}

% Your author list ***DOES NOT*** go here!
% is goes below where you are instructed to insert it...

\author{CLEO Collaboration}
\date{\today}

\maketitle
\tighten

\begin{abstract}
The decay $\tau^- \to K^{*-}\eta \nu_{\tau}$ has been observed with the CLEO~II
detector. The $K^{*-}$ is reconstructed in two decay channels,
$\KSP$ and $\KPZ$. The  $\eta$ 
is reconstructed from the decay $\eta \to \GG$.
The measured branching fraction is
$\mathcal{B}$$(\tau^- \to K^{*-}\eta \nu_{\tau})=
(2.9 \pm 0.8 \pm 0.4 ) \times 10^{-4}$.
We also measure the inclusive branching fractions without requiring 
the $K^*$ resonance,
$\mathcal{B}$$(\tau^- \to K_S \pi^- \eta\nu_{\tau}) =(1.10\pm0.35\pm0.11)
\times 10^{-4} $ and  $\mathcal{B}$$(\tau^- \to K^- \pi^0 \eta\nu_{\tau}) 
=(1.77\pm0.56\pm0.71) \times 10^{-4} $. The results indicate that 
the $K^{*-}$ resonance dominates the  $K_S \pi^-$ mass spectrum.

\end{abstract}  
\newpage

{
\renewcommand{\thefootnote}{\fnsymbol{footnote}}

% Insert author and address list here
\begin{center}
M.~Bishai,$^{1}$ S.~Chen,$^{1}$ J.~Fast,$^{1}$
J.~W.~Hinson,$^{1}$ N.~Menon,$^{1}$ D.~H.~Miller,$^{1}$
E.~I.~Shibata,$^{1}$ I.~P.~J.~Shipsey,$^{1}$
S.~Glenn,$^{2}$ Y.~Kwon,$^{2,}$
% ...
\end{center}
 
\small
\begin{center}
$^{1}${Purdue University, West Lafayette, Indiana 47907}\\
$^{2}${University of Rochester, Rochester, New York 14627}\\
% ...
\end{center}

\setcounter{footnote}{0}
}
\newpage

    The study of the hadronic decays of the $\tau$
lepton is important for a better understanding of
the weak hadronic current and its symmetries.  The decays
involving an $\eta$ meson are associated with the
Wess-Zumino-Witten anomaly~\cite{WZW} and are rare.
The first such decay, $\tau^- \to \pi^- \pi^0 \eta \nu_{\tau}$,
was observed by CLEO in 1992 \cite{Artuso} and subsequently by 
ALEPH  \cite{aleph}. 
More recently, CLEO has measured the branching fractions of two other
decays~\cite{conj}, $\mathcal{B}$$(\tau^- \to K^- \eta \nu_{\tau}) = (2.6 \pm 0.5 \pm 0.5)\times 10^{-4}$~\cite{keta} and 
$\mathcal{B}$$(\tau^- \to (3h)^- \eta \nu_{\tau}) =
(3.5^{+0.7}_{-0.6} \pm 0.7)\times 10^{-4}$~\cite{Shelkov}, 
 where $h = \pi$ or $K$. 
 Both measurements are two orders of magnitude higher
than the predictions by Pich~\cite{Pich} based on chiral
perturbation theory.  However, the recent calculation by
Li~\cite{Li} using an effective chiral theory in the limit
of chiral symmetry is in good agreement with these results.
In the calculation, the
former decay proceeds through the vector current
with $K^*$ dominant and the latter decay proceeds
through the axial-vector current with $a_1$ dominant.
For the decay $\tau^- \to (K \pi)^- \eta \nu_{\tau}$, 
Pich predicts that the $K^*$ enhancement in the $K \pi$ system with 
$\mathcal{B}$$(\tau^- \to  K^- \pi^0 \eta \nu_{\tau}) 
\sim 8.8 \times  10^{-6}$ 
and $\mathcal{B}$$(\tau^- \to \pi^- \bar{K^0} \eta \nu_{\tau}) 
\sim 2.2 \times  10^{-5}$
while Li predicts that the $K_1$ axial-vector current is dominant with 
$\mathcal{B}$$(\tau^- \to K^{*-} \eta \nu_{\tau}) = 1.01 \times  10^{-4}$.  
 In this Letter, we report a first measurement of the decay  
$\tau^- \to K^{*-} \eta \nu_{\tau}$,  $\KSPP$  and  $\KPZ$.
We also measure the inclusive branching fractions without requiring 
the $K^*$ resonance. 
 
 The data used in this analysis have been collected from $e^+e^-$
collisions at a center-of-mass energy of $E_{cm}=10.6$~GeV with
the CLEO II  detector at the Cornell Electron Storage Ring 
(CESR).
 The total integrated luminosity of the sample is 4.7 fb$^{-1}$,
corresponding to the production of $4.3 \times 10^{6}$ $\tau$ pairs.
The CLEO II detector has been described in detail elsewhere~\cite{Kub}.

We select $\tau^+\tau^-$ events in which one charged particle from
the tag $\tau$ decay 
is recoiling against one or three charged particles of the signal decay. The 
candidate events must therefore have two or four 
charged tracks and zero net charge. 
To reject beam-gas events, we require that the distance of closest 
approach to the $e^+e^-$ interaction point 
of the non-$K_S$ candidate tracks be within 0.5 cm (5 cm) transverse to 
 (along) the beam  direction.
Each event is divided into two hemispheres (tag vs. signal) using the plane
perpendicular to the thrust axis~\cite{thr}, calculated from 
 both charged tracks and photons.
Photons are defined as energy clusters in the calorimeter of at least 
60 MeV in the barrel, $|\cos\theta|<0.80$, and 100 MeV in the 
end cap, $0.80<|\cos\theta|<0.95$, where $\theta$ is the polar angle
with respect to the beam axis.
There must be two or more photons in the barrel for  the signal hemisphere.
However, if there are more than two (four) photons 
with an energy above 100 MeV, including the end cap, 
the event is rejected in the $\tau^- \to K_S \pi^- \eta \nu_{\tau}$ 
($\tau^- \to K^- \pi^0 \eta \nu_{\tau}$) analysis.
 The opening angle between the total momentum vectors of the decay products 
of the two $\tau$ leptons must be greater than $120^{\circ}$.  
The tag hemisphere must contain only one charged particle, and its momentum 
must be greater than 0.5 GeV/c. The hemisphere may not contain
more than three energetic photons ($E > 100$~MeV).
In the case of two or more photons,
there must be at least one $\pi^0$ candidate reconstructed,
$|M_{\GG} - M_{\pi^0}| < 20$~MeV/c$^2$ ($\sim 3 \sigma$). 
The hadronic background is suppressed by a requirement that the total 
invariant mass of the particles in each  hemisphere 
be less than the $\tau$ mass, $M < 1.78$~GeV/c$^2$. 
Two-photon, Bhabha, and hadronic events are suppressed by the requirements 
on the total visible energy, 
$0.25<E_{tot}/E_{cm}<0.85$, and on the measured net transverse momentum of
the event, $p_{\perp}>0.3$ GeV/c. 
All charged particles and photons are included 
in the calculation of these kinematic variables.

  Particle identification for  the $\tau^- \to  K^- \pi^0 \eta \nu_{\tau}$ 
decay  is based on a confidence  level ratio which is constructed from 
the confidence levels for $\pi$ and $K$ hypotheses~\cite{keta}, 
$CL_\pi$ and $CL_K$. The confidence level ratio for $K$ is  
$R_K=CL_K/(CL_{\pi}+CL_K)$, and similarly for $\pi$ ($R_{\pi}=1-R_K$).
The confidence level is computed from 
the $\chi^2$ probability for a particle hypothesis using
a combination of the time of flight and drift chamber ($dE/dx$) information. 
 
  Candidate $K_S$ mesons are reconstructed using pairs of oppositely 
charged tracks with vertices separated from the primary interaction point 
by at least 10~mm in the plane transverse to the beam.
The $\pi^+\pi^-$ invariant mass is required to be within 15~MeV/c$^2$ 
 ($\sim 3\sigma$) of the $K_S$ mass. 

The $\eta$ mesons are reconstructed with photons in the barrel
using the $\GG$  decay channel.
Each photon must have an energy above 150 MeV and  a lateral profile
of energy deposition consistent with that expected of a photon.
 In addition,  we do not use the  fragments of
a nearby large shower.
The photon may not combine  with any other photon to form a $\pi^0$
candidate.

\begin{figure}[htbp]
\centerline{\epsfig{file=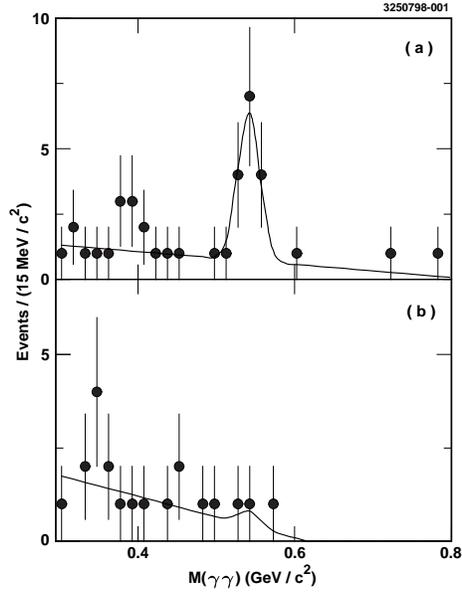,width=2.45in}}
\caption{The invariant mass spectrum of the photon pairs 
in the signal hemisphere containing a $K_S$ 
candidate. The $K_S \pi^-$ invariant mass is required to be in the
$K^{*-}$ signal band in (a) and in the $K^{*-}$ sideband in (b).
The curves show fits to the data.}
\label{fig:mgg}
\end{figure}

 For the  $\tau^- \to K^{*-} \eta \nu_{\tau} \to   K_S \pi^- \eta \nu_{\tau}$ 
analysis, events with three  charged particles in the signal
hemisphere were selected. Figure \ref{fig:mgg} shows the invariant mass 
spectra of  two photons accompanying the $K_S$  candidate, 
with the requirements that the $K_S \pi^-$ mass
be in the $K^{*-}$ signal band (0.81 - 0.97 GeV/c$^2$) or 
sidebands (0.70 - 0.78, 1.00 - 1.08 GeV/c$^2$). 
An $\eta$ signal is observed in the $K^{*-}$ signal region, and there is
no indication of a signal in the sideband region.
The curves show fits to the data using a Gaussian signal and  
a  linear background. The width of the Gaussian is constrained 
to the Monte Carlo expectation, $\sigma = 14$~MeV/c$^2$.
The fit shown in Fig.~\ref{fig:mgg}(a) yields a signal of
$13.3 \pm 3.9$ events. The $\eta$ yield in the $K^{*-}$ 
sidebands is $1.0^{+1.7}_{-1.0}$ events. We have therefore observed for 
the first time the decay $\tau^- \to K^{*-} \eta \nu_{\tau}$.

\begin{figure}[htbp]
\centerline{\epsfig{file=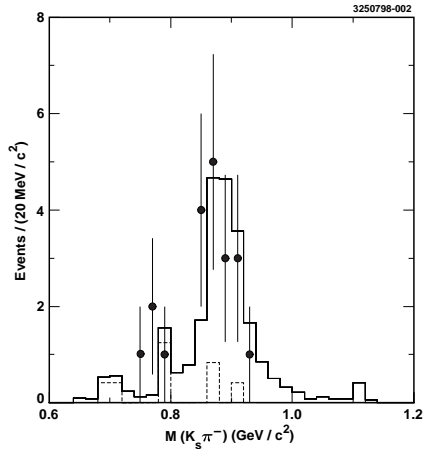,width=2.25in}}
\caption{The invariant mass spectrum of the $K_S  \pi^-$ system 
in the signal hemisphere containing an $\eta$ candidate. The histograms show
the Monte Carlo expectation, including the hadronic background (dotted).}
\label{fig:mksp}
\end{figure}

As a check of the validity of the signal
for $\tau^- \to K^{*-} \eta \nu_{\tau}$, we show
 the invariant mass spectrum of the $K_S  \pi^-$ system for events
with an $\eta$  candidate ($|M_{\gamma\gamma} - M_{\eta}| < 45$~MeV/c$^2$)
in  Fig.~\ref{fig:mksp}. 
A clear $K^{*-}$ signal is observed. 

\begin{figure}[htbp]
\centerline{\epsfig{file=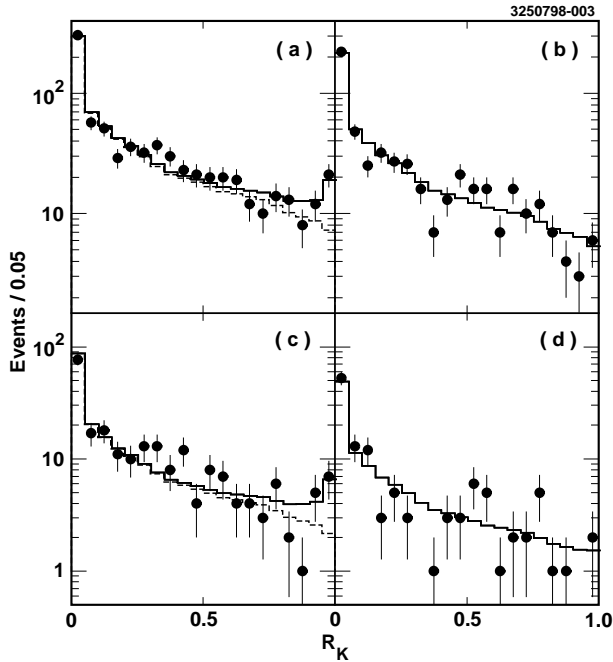,width=3.2in}}
\caption{The $R_K$ spectrum of the charged particle in the signal hemisphere
containing the $\eta$ candidate.  
The invariant mass of the two photons accompanying the charged particle
and $\pi^0$ candidate is required to be in the $\eta$ signal band
in (a) and in the $\eta$ sideband in (b).
Assuming the charged particle is a kaon,
the $K^- \pi^0$ mass is required to be in the $K^*$  band
in (c) and (d). The histograms show fits to the data with the
pions contribution indicated by the dashed histograms.}
\label{fig:plotpr}
\end{figure}

 For the $\tau^- \to K^{*-} \eta \nu_{\tau} \to K^- \pi^0 \eta \nu_{\tau}$ 
analysis, we select events with the signal hemisphere containing
a charged particle, a  $\pi^0$ candidate reconstructed 
using barrel photons plus two other barrel photons.
The $R_K$ distributions for the charged particle in the signal hemisphere
is shown in Fig.~\ref{fig:plotpr}.
The invariant mass of the two photons accompanying the charge particle
and $\pi^0$ candidate is required to be (a) in the $\eta$ signal band
(0.50 - 0.59 GeV/c$^2$), and (b) in the $\eta$ signal sideband
(0.440 - 0.485, 0.605 - 0.650 GeV/c$^2$).
Figures~\ref{fig:plotpr}(c) and ~\ref{fig:plotpr}(d) 
show the corresponding distributions for the case in which
the $K^- \pi^0$ mass is in the $K^{*-}$ signal band with the assumption
that the charged particle is a kaon.
There are enhancements 
at $R_K =$ 0 and, in (a) and (c), 1.0, as expected from the 
decays $\tau^- \to \pi^- \pi^0 \eta \nu_\tau $ and
$\tau^- \to K^- \pi^0 \eta \nu_\tau $ respectively. The histograms show 
fits to the data using the Monte Carlo expectation for $R_K$ spectra for 
these two decays and the migration from other $\tau$ decays. 
The fit results on the number of events with a kaon accompanying the $\eta$
candidate are summarized in Table~\ref{table:rkres}.

 The detection efficiencies for the candidate events and background
from hadronic events are calculated with a Monte Carlo simulation.
The KORALB program \cite{Koral} is used to generate $\tau^+\tau^-$ pairs
and the Lund program \cite{Lund} for hadronic events. 
The signal decays are modeled by phase space assuming 
a $V-A$ weak interaction. The detector response is simulated using the GEANT 
program \cite{Geant}. The identification and misidentification
efficiencies of pions and kaons are calibrated as a function of momentum by 
comparing the efficiencies measured from samples of pions and 
kaons from the decays $D^{*+} \to D^0 \pi^+ \to K^- \pi^+ \pi^+$ 
and $K_S \to \pi^+ \pi^-$ with the hadronic Monte Carlo expectations. 
In the estimation of the hadronic background, the $\eta$
multiplicity in the hadronic Monte Carlo program has been
normalized to produce the observed multiplicity in events
with the invariant mass of one of the hemispheres greater
than $M_\tau$.
Two-photon interactions are estimated to be a negligible
source of background~\cite{keta}.

%\begin{center}
%\begin{minipage}{15cm}
\begin{table}[hp]
\caption{Number of events with a charged 
kaon from the fits of Fig.~\ref{fig:plotpr}
together with the  $\chi^2$ per degree of freedom.}
\begin{tabular}{lccc} 
Requirements & Fig.~\ref{fig:plotpr} & $N_K$ &  $\chi^2$/DOF  \\\hline
$\eta$ signal band & (a) & $36.4 \pm 11.4$ & 23/18 \\
$\eta$ sideband & (b) & $0.0^{+6.7}_{-0.0}$ & 35/18  \\
$\eta$ signal band, $K^*$ region & (c) &  
   $11.7 \pm 5.6$ & 25/18  \\
$\eta$ sideband, $K^*$ region & (d) &   
  $1.0^{+3.1}_{-1.0} $ & 21/15  \\
\end{tabular}
\label{table:rkres}
\end{table}
%\end{minipage}
%\end{center}

 The signals, backgrounds, and detection efficiencies are summarized in 
Tables~\ref{table:e1} and  \ref{table:erk}.
In calculating the detection efficiencies and backgrounds in the 
$\KPZ$ analysis, the Monte Carlo predictions have been corrected for the
appropriate momentum-dependent identification and misidentification
efficiency scaling factors.
The branching fraction for $\tau^- \to (3h)^- \eta \nu_{\tau}$~\cite{Shelkov}
is used to estimate the feeddown
in the $\tau^- \to K_S \pi^- \eta \nu_{\tau}$ analysis.   
 
%\begin{center}
%\begin{minipage}{15cm}
\begin{table}[htbp] 
\caption{ Summary of signals, backgrounds, detection 
efficiencies, and branching fractions for the decay
$\tau^- \to K_S \pi^- \eta \nu_\tau$. 
All errors are statistical.}
\begin{tabular}{lcc}
$K^*$ requirement & Yes & No \\ \hline 
  Signal  & $ 13.3 \pm 3.9 $ & $ 15.1 \pm 4.5 $\\ 
  Signal ($K^*$ sideband) & $ 1.0^{+1.7}_{-1.0} $ & -  \\ 
 $ q \bar{q}$ & $ 0.0^{+1.4}_{-0.0} $ & $ 0.5^{+0.7}_{-0.5} $ \\ 
 $3h \eta$ & $0.4 \pm 0.2$ & $0.6 \pm 0.3$   \\ 
 $3h \eta$ ($K^*$ sideband) & $0.2 \pm 0.2$ & -   \\ 
Eff. (\%) & $4.4 \pm 0.1$ & $5.5 \pm 0.1$  \\ \hline  
 $B$ $(10^{-4})$ & $1.18 \pm 0.38$ & $1.10 \pm 0.35$  \\ 
\end{tabular}
\label{table:e1}
\end{table}
%\end{minipage}
%\end{center}

%\begin{center}
%\begin{minipage}{15cm}
\begin{table}[htbp] 
\caption{ Summary of signals, backgrounds, detection efficiencies,
 and branching fractions for the 
decay $\tau^- \to K^- \pi^0 \eta \nu_\tau$. All errors are statistical.}
\begin{tabular}{lcc}
  $K^*$ requirement & \mbox{Yes} & \mbox{No} \\ \hline 
  $\eta$ band & $ 11.7 \pm 5.6 $ & $36.4 \pm 11.4$  \\ 
  $\eta$ sideband & $1.0^{+3.1}_{-1.0}$ & $0.0^{+6.7}_{-0.0}$ \\ 
  $ q \bar{q}$ & $ < 3.5 @90\% CL$ & $0^{+2}_{-0}$ \\ 
  $K^- \eta$ & - & $0.4 \pm 0.1$ \\ 
  Eff. (\%)  & $4.6 \pm 0.1$  & $6.0 \pm 0.1$ \\ \hline 
$B$ $(10^{-4})$
 & $0.69 \pm 0.36$ & $1.77 \pm 0.56$  \\
\end{tabular}
\label{table:erk}
\end{table}
%\end{minipage}
%\end{center}

%\begin{center}
%\begin{minipage}{15cm}
\begin{table}[htbp]
\caption{Summary of systematic errors~(\%).}
\begin{tabular}{lcc}
    & $K_S \pi^-\eta\nu_{\tau}$ & $K^- \pi^0\eta\nu_{\tau}$ \\ \hline
$N_{\tau \tau} $            &   1.4   & 1.4     \\ 
$\mathcal{B}$$(\eta \to \GG)$           &   0.8   & 0.8     \\ 
$\mathcal{B}$$(\tau^- \to (3h)^- \eta\nu_\tau)$ &  2    & -     \\
Hadronic background            &    5    & 6     \\
$\eta$ sideband subtraction &  -  & 29     \\
Fit                         &    5    & 27    \\
$K_S$ detection eff.~\cite{ksref} & 2  & - \\
Acceptance                  &  3  & 3      \\
Decay model                 &  4  & 4      \\
MC statistics               &  3  & 2      \\ \hline
Total                       &  10 & 40     \\ 
\end{tabular}
\label{table:ss}
\end{table}
%\end{minipage}
%\end{center}

There are several sources of systematic errors as shown in
Table~\ref{table:ss}.  These include
the uncertainties in the number of $\tau^+\tau^-$ events produced, 
branching fractions, background subtraction, 
fitting procedure, $K_S$ detection efficiency, acceptance calculation, 
decay modeling, as well as the uncertainty due to limited Monte Carlo 
statistics. 
The uncertainty in the $R_K$ spectrum at $R_K = 1$ for pions (tail)  and 
kaons (peak)  is a major source of the systematic error in 
the $R_K$ fitting analysis. 
The $R_K$ kaon peak  depends on the momentum distribution 
which is different for $\tau^- \to K^{*-} \eta \nu_{\tau}$ and non-resonant
$\tau^- \to K^{-} \pi^0 \eta \nu_{\tau}$ decays.
The differences are taken as the systematic error estimate.
The systematic error in the acceptance calculation includes the uncertainties 
in the simulation of the tracking, photon detection and veto efficiencies.
The acceptance depends also on the decay model; the corresponding
systematic error is estimated 
by comparing the detection efficiencies for the decays 
$\tau^- \to K_S \pi^- \eta \nu_\tau$ and $\tau^- \to K^- \pi^0 \eta \nu_\tau$
with and without the $K^*$ resonance.

The branching fractions for $\tau^- \to K^{*-} \eta \nu_{\tau}$,
$\KSPP$  and  $\KPZ$ are extracted after correcting for 
backgrounds and detection efficiencies. The results are

\centerline{$\mathcal{B}$$(\tau^- \to K^{*-}\eta\nu_{\tau}) \times \mathcal{B}$$(\KSPP) = (1.18\pm0.38\pm0.12) \times 10^{-4} $,} 

\centerline{$\mathcal{B}$$(\tau^- \to K^{*-}\eta\nu_{\tau}) \times \mathcal{B}$$(\KPZ) = (0.69\pm0.36\pm0.28)\times 10^{-4} $.} 
%\begin{minipage}{\textwidth}
%\hspace{0.4 in} $\mathcal{B}$$(\tau^- \to K^{*-}\eta\nu_{\tau}) \times \mathcal{B}$$(\KSPP)$
%
%\hspace{2.8 in} $=(1.18\pm0.38\pm0.12) \times 10^{-4} $,
%
%\hspace{0.4 in} $\mathcal{B}$$(\tau^- \to K^{*-}\eta\nu_{\tau}) \times \mathcal{B}$$(\KPZ)$
% 
%\hspace{2.8 in} $=(0.69\pm0.36\pm0.28)\times 10^{-4} $.
%\end{minipage}

\noindent Combining these results with the isospin requirement 
$\mathcal{B}$$(\KSPP) =$ $\mathcal{B}$$(\KPZ)= 1/3 $ yields

\centerline{$\mathcal{B}$$(\tau^- \to K^{*-}\eta\nu_{\tau})= 
(2.90 \pm 0.80 \pm 0.42) \times 10^{-4}$.}

The inclusive measurements without the $K^*$ resonance requirement are

\centerline{ 
$\mathcal{B}$$(\tau^- \to K_S \pi^- \eta\nu_{\tau})  = (1.10\pm0.35\pm0.11)
\times 10^{-4} $,}

\centerline{ 
$\mathcal{B}$$(\tau^- \to K^- \pi^0 \eta\nu_{\tau})  = (1.77\pm0.56\pm0.71)
\times 10^{-4} $,}

\noindent
where the first error is statistical and the second systematic.
The inclusive results are in reasonable agreement with the 
measurements requiring the $K^*$ resonance.   

In summary we have measured for the first time the branching fraction of 
$\tau^- \to K^{*-} \eta \nu_{\tau}$. 
The result is somewhat higher than the theoretical prediction by Li~\cite{Li}
($1.01 \times 10^{-4}$).
We also measure the inclusive branching fractions without requiring 
the $K^*$ resonance. The measurements are significantly higher 
than the theoretical predictions by Pich~\cite{Pich}.
The results for the $\tau^- \to K_S \pi^- \eta \nu_{\tau}$ mode indicate that
the $K^{*-}$ resonance dominates the  $K_S \pi^-$ mass spectrum.

We gratefully acknowledge the effort of the CESR staff in providing us with
excellent luminosity and running conditions.
This work was supported by
the National Science Foundation,
the U.S. Department of Energy,
Research Corporation,
the Natural Sciences and Engineering Research Council of Canada,
the A.P. Sloan Foundation,
the Swiss National Science Foundation,
and the Alexander von Humboldt Stiftung.


\begin{thebibliography}{99}

\bibitem{WZW} J. Wess and B. Zumino, Phys. Lett. {\bf B37}, 95 (1971);
               E. Witten, Nucl. Phys. {\bf B223}, 422 (1983).  

\bibitem{Artuso} CLEO Collaboration, M. Artuso {\it et al.},
     Phys. Rev. Lett. {\bf 69}, 3278 (1992).    

\bibitem{aleph} ALEPH Collaboration, D. Buskulic {\it et al.},
     Z. Phys. {\bf C74}, 263 (1997).    

\bibitem{conj} Charge conjugation is implied throughout the paper.

\bibitem{keta} CLEO Collaboration, J. Bartelt {\it et al.}, 
Phys. Rev. Lett. {\bf 76}, 4119 (1996). 

\bibitem{Shelkov} CLEO Collaboration, T.Bergfeld {\it et al.},
  Phys. Rev. Lett. {\bf 79}, 2406 (1997).

\bibitem{Pich}A. Pich, Phys. Lett. {\bf B196}, 561 (1987).

\bibitem{Li} B. A. Li, Phys. Rev. {\bf D55}, 1436 (1997);
                         Phys. Rev. {\bf D57}, 1790 (1998).

\bibitem{Kub} CLEO Collaboration, Y. Kubota {\it et al.}, 
Nucl. Instrum. Methods {\bf A320}, 66 (1992). 

\bibitem{thr} E. Farhi, Phys. Rev. Lett. {\bf 39}, 1587 (1977).

\bibitem{Koral}S. Jadach and Z. Was, Comput. Phys. Commun. {\bf 36}, 
 191 (1985); 
{\bf 64}, 267 (1991); S. Jadach, J.H. Kuhn, and Z. Was, {\it ibid.}
{\bf 64}, 275 (1991).
 
\bibitem{Lund}T. Sj\"{o}strand and M. Bengtsson, Comput. Phys. Commun.
{\bf 43}, 367 (1987).

\bibitem{Geant}R. Brun {\it et al.}, CERN Report No. CERN-DD/EE/84-1, 1987
(unpublished).  

\bibitem{PDG}
Review of Particle Properties, R.M. Barnett {\it et al.}, 
 Phys. Rev. {\bf D54} (1996).

\bibitem{ksref} CLEO Collaboration, M. Bishai {\it et al.}, 
Phys. Rev. Lett. {\bf 78}, 3261 (1997).

\end{thebibliography}
\end{document}